\documentstyle[aps,prl,epsf,rotate]{revtex}
\begin{document}
%\draft
%\preprint{}
%\input psfig

\twocolumn[\hsize\textwidth\columnwidth\hsize\csname     
@twocolumnfalse\endcsname

\title{Fractional Quantum Hall Effect of Composite Fermions}

\author{W. Pan$^{1,2}$, H.L. Stormer$^{3,4}$, D.C. Tsui$^1$, L.N.
Pfeiffer$^4$, K.W. Baldwin$^4$, and K.W. West$^4$}
\address{$^1$Department of Electrical Engineering, Princeton
University,
Princeton, New Jersey 08544}
\address{$^2$National High Magnetic Field Laboratory,
Tallahassee, Florida 32310}
\address{$^3$Department of Physics
and Department of Applied Physics, Columbia University, New
York,
New York 10027}
\address{$^4$Bell Labs, Lucent Technologies, Murray Hill, New
Jersey 07974}

\date{\today}
\maketitle

\begin{abstract}

In a GaAs/AlGaAs quantum well of density $1 \times 10^{11}$ cm$^{-2}$ 
we observed
a fractional quantum Hall effect at $\nu=4/11$ and 5/13,
and weaker states at $\nu=6/17, 4/13, 5/17$, and 7/11. These sequences
of fractions do not fit into the standard series of integral quantum
Hall effects (IQHE) of composite fermions (CF) at $\nu = p/(2mp\pm1)$.
They rather can be regarded as the FQHE of CFs attesting to residual
interactions between these composite particles. In tilted magnetic
fields the $\nu=4/11$ state remains unchanged, strongly suggesting it to
be spin-polarized. The weak $\nu=7/11$ state vanishes quickly with tilt.

\end{abstract}

%\vskip2pc
\vskip2pc]

The composite fermion (CF) model \cite{jain:prl89,fradkin:prb91,zhang:prb92,hlr:prb93,cfbook1,cfbook2} has been very successful in
providing a rationale for the observed sequences of principal
fractional quantum Hall effect (FQHE) states at Landau level fillings
$\nu=p/(2mp\pm1)$ ($p$, $m$ = 1, 2, 3, $\cdot \cdot \cdot$) around major even-denominator
fractions, $\nu=1/2m$. In this model, the dominating electron-electron interaction is
very effectively incorporated into the carriers by transforming them
into new particles, $^{2m}$CFs, by virtue of the attachment of an even
number, 2$m$, of magnetic flux quanta. As a consequence CFs can be
treated as independent particles in an effective magnetic field, $B_{eff}$,
which is reduced from the external field, $B$, by the density of the
attached magnetic flux. As $B_{eff}$ deviates from zero, Landau levels of
CFs develop, giving rise to an integral quantum Hall effect (IQHE) of
these flux-transformed, non-interacting composite particles. This IQHE
of CFs in the effective magnetic field becomes equivalent to the FQHE
of the original, highly interacting electrons exposed to the full
external field. Experiments in the FQHE regime closely follow the
sequence of proposed states according to this model. Prime examples
for the applicability of this model are the sequences of FQHE states
at filling factors $\nu=p/(2p\pm1)$ around $\nu=1/2$ -- made from $^2$CFs --
and $\nu=p/(4p\pm1)$ around $\nu=1/4$ -- made from $^4$CFs --  which match,
when appropriately shifted, the sequence of IQHE states around $B=0$. 

The question arises as to the ultimate validity of the assumption of
vanishing interaction between CFs. Much of the modeling of CF physics
requires a mean field approach, whose exact applicability to the
conditions at hand is doubtful. Residual interactions between CFs may
simply lead to small corrections of the CF properties or -- more
interestingly --  they themselves may create novel electronic states.
Experimentally, distortions of the shape of $R_{xx}$ maxima between
neighboring FQHE states (or IQHE states of CFs) have hinted in the
past towards residual CF-CF interactions \cite{horst:physicae98}. In this paper we present
extensive experimental evidence for the considerable strength of these
interactions by observing the appearance of FQHE states at filling
factors $\nu=4/11$, 5/13, 7/11, 4/13, 6/17 and 5/17, located between the
minima of the primary FQHE sequences. Regarding the primary sequences
as the IQHE of CFs, the new states can be viewed as the FQHE of CFs,
brought about by CF-CF interactions. From angular dependent
measurements we can deduce the spin polarization of the strong $\nu=4/11$
state and find it to be spin polarized in our specimen. Its much
weaker, electron-hole symmetric state at $\nu=7/11$ rapidly disappears
under tilt leaving its spin polarization uncertain.

The sample consists of a 500 \AA-wide modulation-doped GaAs/AlGaAs
quantum well and has a size of about 5mm $\times$ 5mm. 
The well is $\delta$-doped
with silicon from both sides at a distance of $\sim$ 2200 \AA. Electrical
contacts to the two-dimensional electron system (2DES) are
accomplished by rapid thermal annealing of indium beats along the
edge. An electron density of $n \sim 1.0\times10^{11}$ cm$^{-2}$ 
and a mobility of $\mu \sim 
10\times10^6$ cm$^2$/Vs were achieved after illumination of the sample at low
temperatures by a red light-emitting diode. A self-consistent
calculation shows that at this density only one electrical subband is
occupied, consistent with the low-field Shubnikov-de Haas data.
Conventional low frequency ($\sim$ 7Hz) lock-in amplifier techniques were
employed to measure the magnetoresistance $R_{xx}$ and Hall resistance $R_{xy}$.

Fig.1 shows an experimental trace of $R_{xx}$ in the regime of $2/3>\nu>2/7$,
taken at $T \sim 35$ mK. The very low and very high-field data are omitted to
emphasize the field range central to this study. Several outstanding
features are apparent in Fig.1: (1) Very high-denominator FQHE states
at $\nu=10/19$ and $\nu=10/21$ appear around $\nu=1/2$, attesting to the
exceptionally high quality of the specimen. (2) Slight undulations,
previously observed between $\nu=1/3$ and 2/5 \cite{horst:physicae98}, are resolved into
clear $R_{xx}$ minima at $\nu=5/13$, 3/8, 4/11, and 6/17. The minimum at
$\nu=4/11$ is particularly strong. (3) $R_{xx}$ minima are even observed at
fractions $\nu=4/13$, 3/10, and 5/17 between $\nu=1/3$ and 2/7.  Between
$\nu=2/3$ and 3/5, a minimum appears at $\nu=7/11$ \cite{goldman:ss90} and a weaker feature
around $\nu=5/8$. (4) A clear slope change is apparent in $R_{xy}$ (dashed
lines) at $\nu=4/11$ and 7/11, although, as in the early stages of many
developing fractions, the quantization value remains poorly defined. 

The sum of new fractions, appearing between traditional FQHE states,
can be understood naively as the FQHE of CFs. In terms of $^{2m}$CF Landau
level filling factor, $\nu_{2m}$, the states at $\nu=4/11$, 5/13, 6/17 reside
between the consecutive minima of the $\nu_2=1$ and $\nu_2=2$ IQHE of $^2$CFs as
counted from $B_{eff} =0$ at $\nu=1/2$. In the same spirit, the $\nu=4/13$ and
5/17 reside between the $\nu_4=1$ and $\nu_4=2$ IQHE of $^4$CFs as counted from
$B_{eff}=0$ at $\nu=1/4$. At $\nu=4/11$ the 
lowest $^2$CF Landau level, $\nu_2=1$, is
completely filled whereas the second $^2$CF Landau level is filled to 1/3
of its $^2$CF capacity. Hence $\nu=4/11$, in terms of electrons, corresponds
to $\nu_2=1+1/3$ in terms of $^2$CFs, which, therefore, can be regarded as
the 1+1/3 FQHE of $^2$CFs. The correspondence of the other fractions is
as follows: $\nu=5/13 \to \nu_2=1+2/3$, $\nu=6/17 \to \nu_2=1+1/5$, $\nu=3/8
\to \nu_2=1+1/2$, $\nu=4/13 \to \nu_4=1+2/3$, $\nu=5/17 \to \nu_4=1+1/3$,
$\nu=3/10 \to \nu_4=1+1/2$, $\nu=7/11 \to \nu_2=1+1/3$, 
$\nu=5/8 \to \nu_2=1+1/2$.
There are also hints in the data for further features between $\nu=2/5$
and $\nu=3/7$ as well as between $\nu=2/9$ and $\nu=1/5$. 
Furthermore, we have observed
similar deformation in $R_{xx}$ traces of other, higher density samples
(not shown) between $\nu=3/5$ and $\nu=4/7$, 
between $\nu=2/3$ and $\nu=5/7$, and 
between $\nu=1+1/3$ and $\nu=1+2/5$ as well as between $\nu=1+2/7$ and
$\nu=1+1/3$ \cite{zhu:unpublished}, 
hinting towards a continuation of FQHE states of CFs to
higher CF Landau levels. 

The relative strength of these features resembles the relative
strength of the electron FQHE and 
the progression of their discovery \cite{fqhe82,qhebook}.
A selfsimilarity seems to be at work in which the pattern of
FQHE features, initially observed in electrons, is now observed in CFs
and may progress further to higher-order CFs in yet lower disorder
samples. Of course, one may already regard the sequence of FQHE state
at $\nu=p/(4p\pm1)$ as the FQHE of $^2$CFs, since their lowest $^2$CF Landau
level is only partially occupied. However, the situation is equally
well, and more naturally, described as the IQHE of $^4$CF, emanating from
$\nu=1/4$ \cite{hlr:prb93}. Instead of two flux quanta, 
each electron is now carrying
four flux quanta. 

For the new sequences the hypothetical flux attachment process
would be much more intricate. For example the $\nu=4/11$ state is created
by the following mental sequence. Two flux quanta attach themselves to
each electron forming $^2$CFs, which, at $\nu=1/2$, form a fermi sea with
$B_{eff}=0$. At $\nu=1/3$, the lowest $^2$CF Landau level created by the now
finite $B_{eff}$ has reached a degeneracy sufficient to accept all $^2$CFs and
the $\nu_2=1$ IQHE of $^2$CF occurs. As $B_{eff}$ is reduced to 1/2 of its
strength, all $^2$CFs fill exactly two $^2$CF Landau levels and the $\nu_2=2$
IQHE of $^2$CFs occurs (equivalent to $\nu=2/5$). At $\nu=3/8$, equivalent to
$\nu_2=3/2$ the lowest $^2$CF Landau level is totally occupied, whereas the
second $^2$CF Landau level is occupied only to 1/2 of its $^2$CF capacity,
assuming total spin-polarization. At this stage, two flux quanta attach
themselves to the $^2$CFs in the higher $^2$CF Landau level. The lower one,
being completely full, can be ignored. In total, 1/3 of all CFs have
become $^4$CFs, whereas 2/3 remained $^2$CFs \cite{park:prb00}. This is a rather
intricate situation, in which every 3 electron carry 8 flux quanta. It
exactly cancels out the external $B$-field 
at $\nu=3/8$ and $B_{eff}=0$, again \cite{hlr:prb93}.
As $B$ moves toward $\nu=4/11$ the rising $B_{eff}$ creates $^4$CF Landau
levels with a degeneracy sufficient to accept all $^4$CFs into the lowest
$^4$CF Landau level and a $\nu_4=1$ IQHE of $^4$CFs occurs. This would be the
$\nu=4/11$ FQHE state.

The other observed fractions can be derived in an equivalent
fashion. The $\nu=6/17$ requires 1/5 $^6$CFs and 4/5 $^2$CFs. The states around
$\nu=3/10$ requires 2/3 $^2$CFs and 1/3 $^4$CFs. In all cases we have assumed
complete spin polarization, as one may expect at such high fields. If
this were indeed the case, then the $\nu=3/8$ (equivalent to $\nu_2=3/2$)
would, in fact, be equivalent to the $\nu=5/2$ electron state, since it
occupies the second $^2$CF Landau level and not the upper spin state of
the lowest Landau level as is the case for electrons. This may explain
the observation of a minimum at $\nu=3/8$. At $\nu=5/2$ electrons show a
FQHE, believed to originate from paired $^2$CFs. At $\nu=3/8$ the $^4$CFs of
the fermi liquid may pair and form a paired state of $^4$CFs \cite{jain:prl01}.
Furthermore, the suppression in $R_{xx}$ observed around $\nu=5/12$ between
$\nu=2/5$ and $\nu=3/7$, equivalent to $\nu=5/2$ would, in fact, be equivalent
to the anisotropic electron state at $\nu=9/2$ \cite{lilly:prl99,du:ssc99}, 
since all spins
are polarized due to the large external $B$-field. 

Beyond the qualitative demonstration of this apparent
selfsimilarity in the FQHE sequences we have also performed
quantitative studies on the stronger of the fractional states. Fig.2
summarizes the $T$-dependences for the states at $\nu=4/11$ and 7/11
states. Three temperature traces are shown in Fig. 2(a) and four in
Fig. 2(b). Unlike the well-developed FQHE states, $R_{xx}$ at exactly
$\nu=4/11$ barely changes with temperature. On the other hand, the
strength of the whole $\nu=4/11$ feature decreases markedly with
increasing temperature. Such a $T$-dependence is reminiscent of the
initial $T$-dependence of many FQHE states. In analogy to earlier
procedures \cite{gammel:prb88,eisenstein:prl88}, 
we deduce the gap energy of the $\nu=4/11$ state of
approximately 30-50 mK from the strength of its minimum. The enormous
change of shape of the data around $\nu=7/11$ leads to an erratic
$T$-dependence and no effective energy scale can be deduced. 

In Fig.3 we address the spin polarization of the strongest
state. At $B=11$ T the Zeeman energy of electrons in GaAs is $\sim$ 3K.
These
energies are vastly larger than the characteristic energies (several
mK) extracted from Fig. 2(a). Already at this point it appears highly
unlikely that $\nu=4/11$ is spin unpolarized. The ground state energy and
Zeeman energy would have to balanced each other closely in order to
show such small gap energies for the states.
To further strengthen this conjecture we
measured $R_{xx}$, {\it in situ}, 
as a function of tilted magnetic field as shown
Fig.~3 for the $\nu=4/11$ minima and several representative angles,
$\theta$ \cite{note1}. 
For an ideal 2DES the correlation energy remains unchanged under
tilt, while the Zeeman energy increases as 1/cos($\theta$), 
rising to $\sim$ 4.5K
at $\theta \sim 42^{\circ}$ for the $\nu=4/11$ state. 
The huge differential increase of
$\sim$ 1.5K over the Zeeman energy at $\theta=0^{\circ}$ 
should move any possible close
balance of correlation energy and Zeeman energy at $\theta=0^{\circ}$, vastly in
favor of the latter at $\theta \sim 42^{\circ}$ and lead to a transition in the spin
polarization. Any such macroscopic change of the spin property between
$\theta=0^{\circ}$ and $\theta \sim 42^{\circ}$ would be visible as a vanishing or strongly reduced
strength of the $\nu=4/11$ state. The data of Fig.3 for tilts up to
$\theta=42.2^{\circ}$ show practically no 
variation in the strength nor the shape
of the $\nu=4/11$ state, from which we infer that no spin transition
occurs. Therefore, the $\nu=4/11$ state in our sample must be spin
polarized for all angles. Even the feature at $\nu=5/13$ shows no angular
dependence. Although this fraction is not very well developed, this
lack of variation suggests this state to be also spin polarized.
 
We also measured the angular dependence of the $\nu=7/11$ state at base
temperature (not shown). In contrast to the $\nu=4/11$ state the features
of the $\nu=7/11$ state change considerably under tilt and the minimum
seems to have disappeared entirely by $\theta=29.5^{\circ}$. 
This points to a spin
transition in the $\nu=7/11$ state and hence a spin unpolarized or, at
least, partially polarized state at $\theta=0^{\circ}$. However, as in previous tilt
data on the $\nu=5/2$ state one cannot rule out that it is an orbital
effect that is destroying the $\nu=7/11$ gap. We therefore refrain from
assigning any spin polarization to the $\nu=7/11$ state at this time. 

The drawing of analogies, such as those presented above, about
the continuations of the CF picture to higher orders in a selfsimilar
scheme is rather simple. However, theoretical calculations as to the
stability of such higher order FQHE states are very difficult, since
they requires many particles to treat the inherent correlations
realistically. Consequently, the theoretical situation regarding such
states remains in flux. The early 
hierarchical model of the FQHE \cite{haldane:prl83,halperin:prl84}
obviously allows for the existence of all of our observed
states, since it covers all odd-denominator fractions \cite{note2}.  However,
several of the newly discovered states are expected not to be stable
\cite{haldane:prl95}.
A paper by Wojs and Quinn studies quasiparticle interactions in
the FQHE regime and compares different hierarchies \cite{quinn:prb00}. It finds the
$\nu=6/17$ state to be possibly stable, but the $\nu=4/11$ and 
$\nu=4/13$ states to be
definitely unstable within their pseudopotential classification
approach and an 8-electron exact diagonalization for zero-thikness
layers. These studies rely on total spin polarization and the apparent
absence of an incompressible state at $\nu=4/11$ persuaded Park and Jain
\cite{park:prb00} to investigate partially 
polarized states at $\nu=4/11$, for which
they find, indeed, a small energy gap. However, both investigation
seem to conflict with our experimental result of the existence of a
spin polarized state at $\nu=4/11$. A very recent preprint by Mandal and
Jain \cite{mandal} tests for the existence of higher order CF states via a new
numerical scheme, which neglects CF Landau level mixing, but can
handle as many as 24 electrons. Surprisingly, this approach generates
no condensed state at any of the higher-order states we observe. The
study concludes that ``the physical mechanism, in which some of the CFs
turn into higher order CFs and condense into new Landau levels to
exhibit a QHE, does not appear to be relevant for fully polarized
electrons''. This would negate a simple, selfsimilar model for CFs as
we have proposed on the basis of the newly observed sequences of FQHE
states. On the one hand, this conflict may arise from some
fundamentally important aspect of the interaction, which has been
omitted in the numerics. On the other hand it may arise from the
neglect of more subtle experimental realities such a finite thickness
of the 2DES or mixing between Landau levels. Yet it appears unlikely
that the stability of all sequences of observed higher-order states
would be a finite-thickness effect. In fact, we have observed the
strong $\nu=4/11$ state and weaker reflection of the other states also in
a triangular well and in square wells of thichnesses from 30~nm to
60~nm. This suggests a large independence of the stability of such
states from the confining potential.

At this stage, the origin of the minima observed in $R_{xx}$ at
$\nu=4/11$, 5/13, 6/17, 4/13, 5/17 and 7/11 is unresolved. Numerical
calculations seem to exclude the existence of incompressible states at
such filling factors for spin polarized systems, at least for
zero-thichness systems. Yet angular dependent measurement on the
$\nu=4/11$ state and the existence of any such state, in spite of large
Zeeman energies compared to their characteristic energies, suggests
all of them to be fully spin polarized. This clearly points to a lack
of our understanding of these new features. Furthermore, the
simple-minded and intuitive continuation of the CF flux attachment
scheme to partially filled CF Landau levels and the analogies that can
be drawn from the sequential observation of the principal FHQE
sequence to the observation of the new fractions seem to work so well
that one is tempted to accept their fundamental appropriateness. The
same naive analogy thinking would also provide a rationale for the
existence of minima at $\nu=3/8$ and $\nu=3/10$, since those would be
equivalent to the $\nu=5/2$ FQHE state. Even the relative strength of the
$\nu=6/17$ state finds its similarity in the state at the $\nu=11/5$. The
aggregate of our experimental observations highly suggests a
continuation of the CF model to higher orders, or, equivalently, the
existence of a FQHE of CFs. Such a selfsimilarity in the sequence of
FQHE states is too appealing to be discarded yet.

We would like to thank the E. Palm, T. Murphy, and S. Hannas for
experimental help. We are grateful to E. Fradkin, F.D.M. Haldane, J.
K. Jain, J. Moore, K. Park, E. Rezayi, and S. Simon for useful
discussions. A portion of this work was performed at the National High
Magnetic Field Laboratory, which is supported by NSF Cooperative
Agreement No. DMR-9527035 and by the State of Florida. D.C.T. and W.P.
are supported by the AFOSR, the DOE, and the NSF. H.L.S. is supported
by DOE and the W.M. Keck Foundation.

\begin{figure}
\caption{
$R_{xx}$ in the regime of $2/3>\nu>2/7$ at $T \sim 35$ mK. 
Major fractions are marked
by arrows. Dashed traces are the Hall resistance $R_{xy}$ around $\nu=7/11$
and $\nu=4/11$.
}
\end{figure}

\begin{figure}
\caption{
(a) $T$-dependence of $R_{xx}$ around $\nu=4/11$. Three temperature traces are
shown: solid line - 35mK; dashed line - 70mK; dotted line - 95mK. 
(b)
$T$-dependence of $R_{xx}$ around $\nu=7/11$. Four temperature traces are shown:
solid line - 40mK; dashed line - 70mK; dotted line - 105mK;
dash-dotted line -185mK.
}
\end{figure}

\begin{figure}
\caption{
$R_{xx}$ between $2/5 > \nu >1/3$ at five selected tilt angles. A total of 18
tilt angles were actually measured. Position of the $\nu=4/11$ state is
marked by arrow.
}
\end{figure}


\begin{references}

\bibitem{jain:prl89}
J.K. Jain, Phys. Rev. Lett. {\bf 63}, 199 (1989).

\bibitem{fradkin:prb91}
A. Lopez and E. Fradkin, Phys. Rev. B {\bf 44}, 5246 (1991).

\bibitem{zhang:prb92}
V. Kalmeyer and S.C. Zhang, Phys. Rev. B {\bf 46}, 9889 (1992).

\bibitem{hlr:prb93}
B.I. Halperin, P.A. Lee, and N.Read, Phys. Rev. B {\bf 47}, 7312
(1993).

\bibitem{cfbook1}
{\it Perspectives in Quantum Hall Effect}, S. Das Sarma and A. Pinczuk
(Eds.), Wiley, New York (1996).

\bibitem{cfbook2}
{\it Composite Fermions: A unified View of the Quantum Hall Regime},
O. Heinonen (Ed.), World Scientific, Singapore (1998).

\bibitem{horst:physicae98}
H.L. Stormer, A.S. Yeh, W. Pan, D.C. Tsui, L.N. Pfeiffer, K.W.
Baldwin, and K.W. West, Physica E {\bf 3}, 38 (1998).

\bibitem{goldman:ss90}
V.J. Goldman and M. Shayegan, Surface Science {\bf 229}, 10 (1990).

\bibitem{zhu:unpublished}
J. Zhu {\it et al}, unpublished.
 
\bibitem{fqhe82}
D.C. Tsui, H.L. Stormer, and A.C. Gossard, Phys. Rev. Lett. {\bf 48},
1559 (1982).

\bibitem{qhebook}
{\it The Quantum Hall Effect}, R.E. Prange and S.M. Girvin (Eds.),
Springer, New York (1990).

\bibitem{park:prb00}
K. Park and J.K. Jain, Phys. Rev. B {\bf 62}, 13274 (2000).

\bibitem{jain:prl01}
S.Y. Lee, V.W. Scarola, and J.K. Jain, Phys. Rev. Lett. {\bf 87},
256803 (2001).

\bibitem{lilly:prl99}
M.P. Lilly, K.B. Cooper, J.P. Eisenstein, L.N. Pfeiffer, and K.W.
West, Phys. Rev. Lett. {\bf 82}, 394 (1999).

\bibitem{du:ssc99}
R.R. Du, D.C. Tsui, H.L. Stormer, L.N. Pfeiffer, K.W. Baldwin, and
K.W. West, Solid State Commun. {\bf 109}, 389 (1999).

\bibitem{gammel:prb88}
P.L. Gammel, D.J. Bishop, J.P. Eisenstein, J.H. English, A.C.
Gossard, R. Ruel, and H.L. Stormer, Phys. Rev. B {\bf 38}, 10128
(1988).

\bibitem{eisenstein:prl88}
J.P. Eisenstein, R.L. Willett, H.L. Stormer, D.C. Tsui, A.C.
Gossard, and J.H. English, Phys. Rev. Lett. {\bf 61}, 997 (1988).

\bibitem{note1}
The data were taken during a different cool-down, which causes the
2DES density to be slightly ($\sim 5\%$) higher.

\bibitem{haldane:prl83}
F.D.M. Haldane, Phys. Rev. Lett. {\bf 51}, 605 (1983).

\bibitem{halperin:prl84}
B.I. Halperin, Phys. Rev. Lett. {\bf 52}, 1583 (1984).

\bibitem{note2}
Possible FQHE at $\nu=3/8$ and 3/10 would require additional
interactions such as at $\nu=5/2$.

\bibitem{haldane:prl95}
F.D.M. Haldane, Phys. Rev. Lett. {\bf 74}, 2090 (1995).

\bibitem{quinn:prb00}
A. Wojs and J.J. Quinn, Phys. Rev. B {\bf 61}, 2846 (2000).

\bibitem{mandal}
S.S. Mandal and J.K. Jain, preprint.

\end{references}
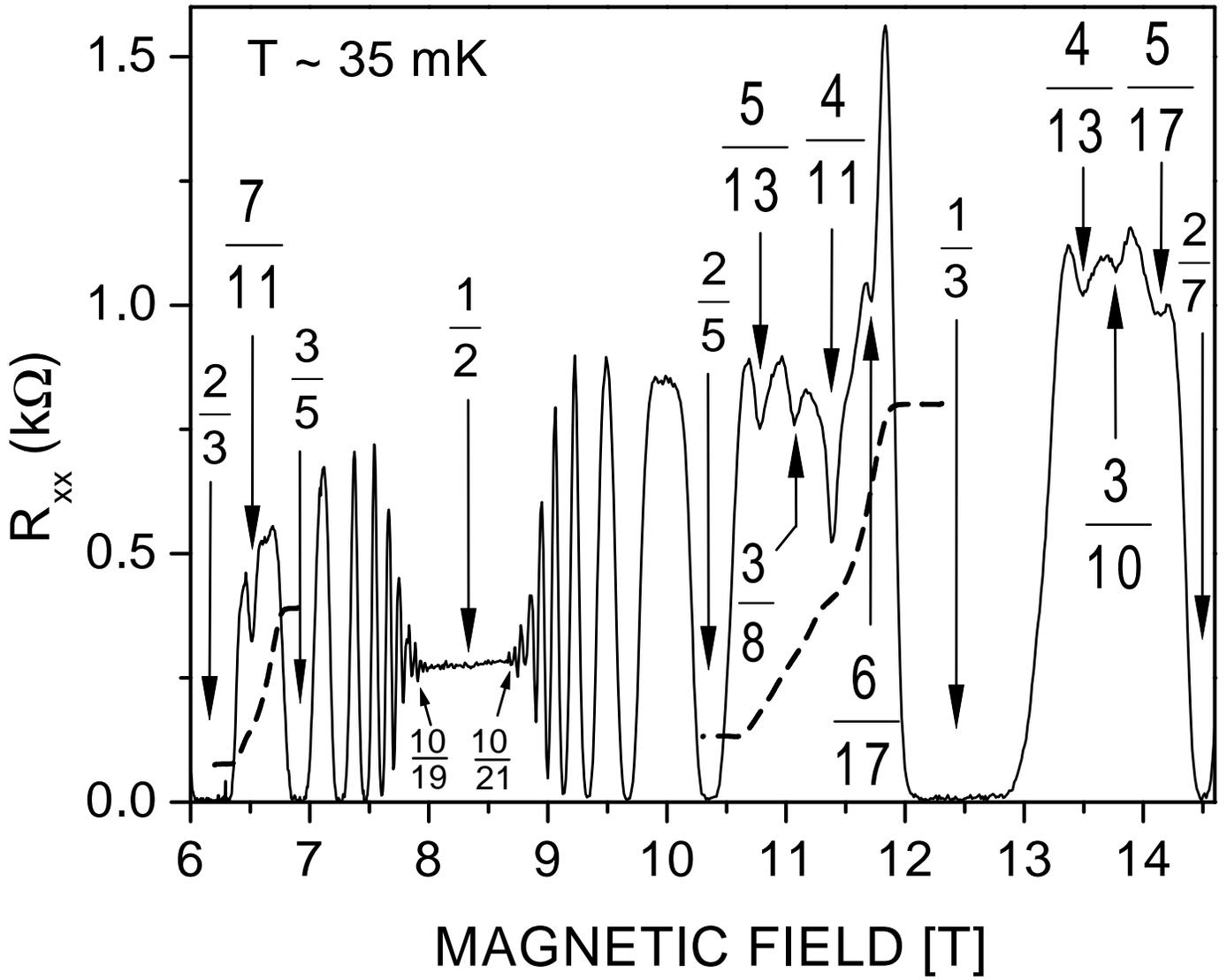
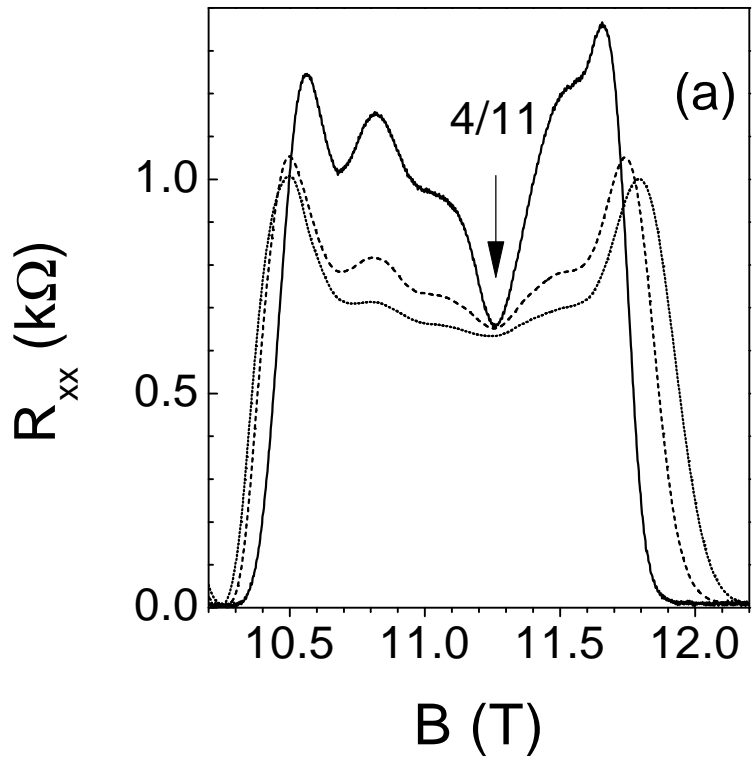
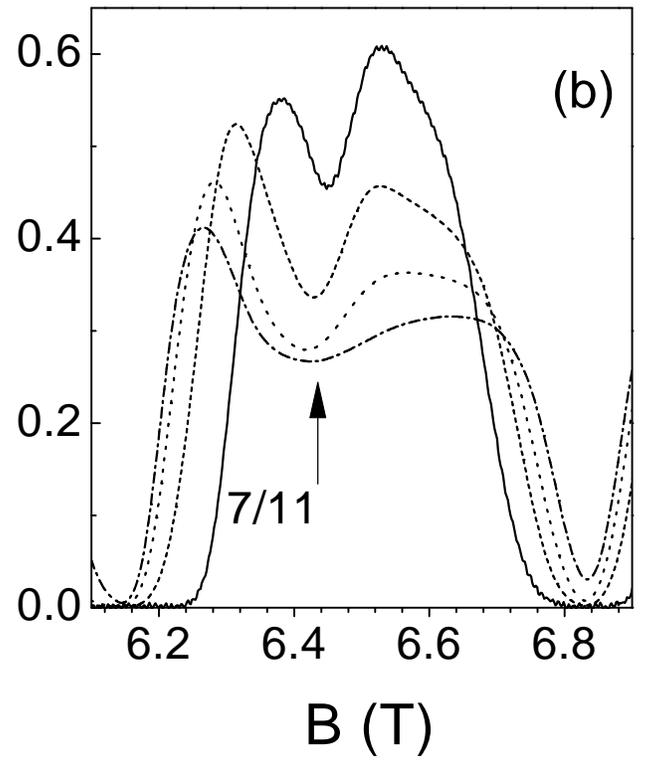
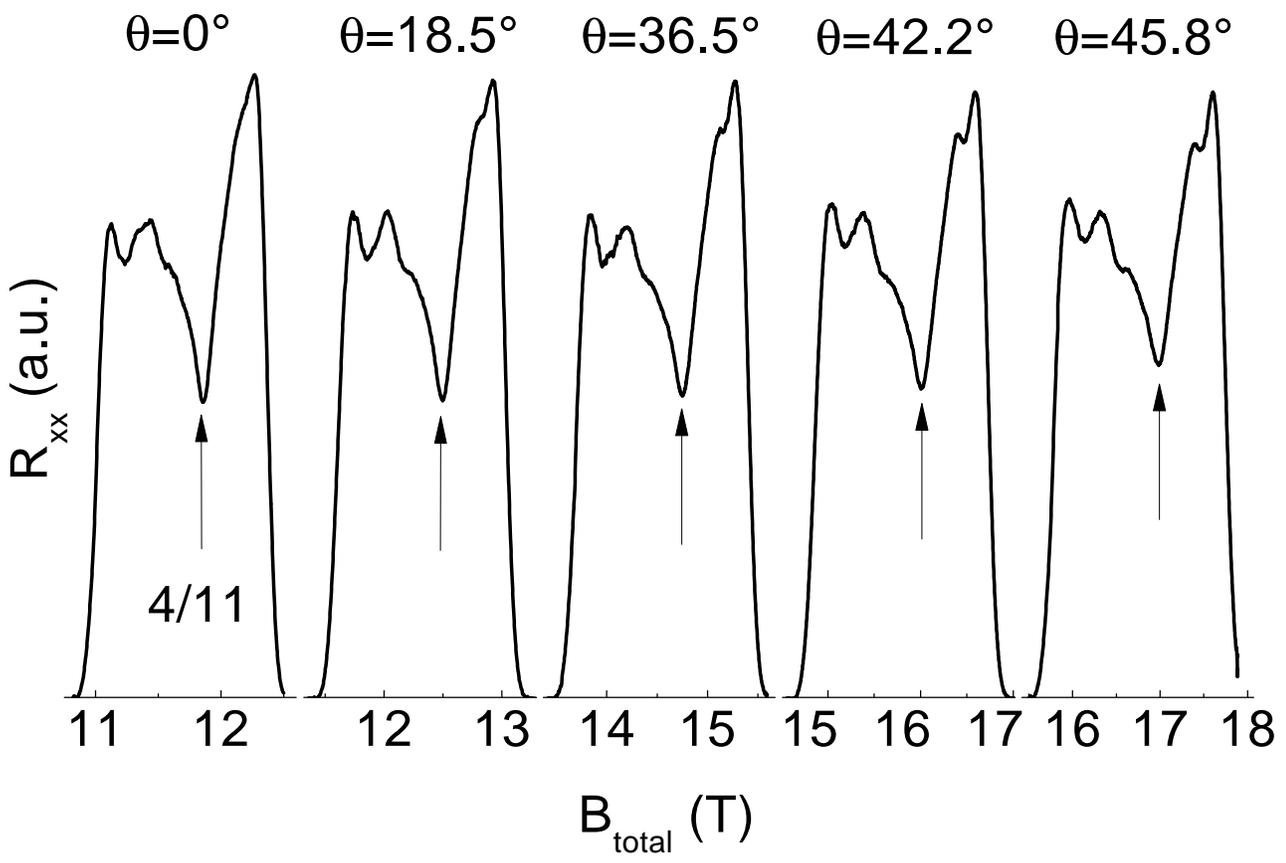
\end{document}